\documentclass[aps,prl,twocolumn,showpacs,superscriptaddress,groupedaddress]{revtex4-1}
\usepackage{graphicx}  
\usepackage{dcolumn}   
\usepackage{bm}        
\usepackage{amssymb}
\begin{document}

\title{Majorana Edge States and Braiding in an Exactly Solvable One-dimensional Spin Model}

\author{Zhao-Yang Dong}
\author{Jian-Xin Li}
\affiliation{Department of Physics and National Laboratory of
Solid State Microstructure, Nanjing University, Nanjing 210093,
China\\
Collaborative Innovation Center of Advanced Microstructures,
Nanjing University, China}

\date{\today}

\begin{abstract}
We derive an exactly solvable one-dimensional (1D) spin model from
the three-band Hubbard model with a strong spin-orbit coupling by
introducing $U (1)$ gauge fields to the isospin states. We find
that it has a topological nontrivial phase characterized by
Majorana end modes which are protected by a new $Z_2$ topological
invariant related to the parity of the lattice sites (odd or even
number of sites) in the spin chain. With the protection of this
$Z_2$ topological invariant, a novel braiding of two Majorana edge
states in this strictly geometric 1D chain is realized. We also discuss the
possible realization of the gauge fields.


\end{abstract}

\maketitle

\emph{Introduction}---Majorana fermions, which are their own
antiparticles, have attracted massive theoretical and experimental
interests recently, mainly because they have a condensed matter
analog in the zero-energy bound states emerging in topological
superconductors (SC). Furthermore, in two dimensions (2D) these
Majorana zero modes manifest non-Abelian braiding statistics.
Exchanging two Majoranas represents a non-commutative operation on the ground states.
Quantum information encoded in this ground state is non-local,
therefore, they have been regarded as an ideal building blocks for
fault-tolerant topological quantum
computation~\cite{Nayak2008,Kitaev2001,Duckheim2011}.

Topological superconductivity was originally recognized in
spin-triplet $p$-wave
SC~\cite{Read2000,Kraus2009,Ivanov2001,Wimmer2010,Fu2008,Potter2011,Alicea2010,Sau2010,Sato2010}.
In the weak-pairing phase of a 2D spinless $p_x+ip_y$
SC, the Bogoliubov-de Gennes quasiparticles bounded to defects
(vortices or sample edges) satisfy the particle-hole symmetry, and
consequently the zero-energy quasiparticles can be identified as
Majorana fermions. However, Majoranas occur in vortices are very
sensitive to disorder and have not been identified experimentally.
A recent promising route to realize topological SC hosting
Majorana fermions is the one-dimensional (1D) semiconductor
nanowires with proximity coupling to an $s$-wave SC in the
presence of strong spin-orbit interaction and an external Zeeman
field~\cite{Sau2010,Oreg2010}. Because it is impossible to
exchange two particles in 1D without bringing them to the same
spatial position in the process, the non-Abelian braiding of two
Majorana fermions in this scheme has been proposed to carry out by
wire networks, such as the T-junction formed by two perpendicular
nanowires~\cite{Alicea2011} or the cross-shaped four
nanowires~\cite{Amorim2014}. But, a delicate gate control over
topological superconducting state poses a significant challenge.

In fact, Majorana fermions can also be realized in some spin-only
system, from as simple as the Ising model to the Kitaev ladder
model~\cite{DeGottardi2011}. For example, the spin-1/2 $XY$ chain
can be mapped exactly to the famous 1D spinless $p$-wave SC
supporting Majorana end modes~\cite{DeGottardi2011}. Furthermore,
it is also proposed that the entangled states in the Heisenberg XY
model can be generated for qubits in quantum
computation~\cite{Wang2001}.

In this paper, we derive a low-energy effective spin model from a
1D Hubbard model with partially filled $t_{2g}$ bands in the presence of a strong spin-orbital
coupling (SOC) by introducing $U (1)$ gauge fields to the isospin
states. We consider a hole resides in the $t_{2g}$ manifold of $xy,xz,yz$ orbitals. The SOC splits this sixfold degenerate states into a half-filled $J_{eff}=1/2$ bands (Kramers doublet) and
completely filled $J_{eff}=3/2$ bands, which gives rise to an isospin $J_{eff}=1/2$ Mott insulator state.
The exchange Hamiltonian for isospin is then obtained by projecting the corresponding superexchange spin-orbital model
in large Hubbard interactions limit on the Kramers doublet. With the proper $U(1)$ gauge fields, the Heisenberg term can be eliminated
and we get an exactly solvable 1D spin model consists of the isospin couplings with only $x$ and $y$ components.
We elaborate that this isospin model has a topological nontrivial phase characterized by
Majorana end modes which are protected by a new $Z_2$ topological
invariant related to the parity of the lattice sites (odd or even
number of sites). With the protection of this
$Z_2$ topological invariant, a novel braiding of two Majorana edge
states in this strict 1D geometric chain is realized. We also give speculations about physical realization
of the introduced gauge fields.

\emph{Effective spin model}---We start from a three-orbital Hubbard model with
one hole in the $t_{2g}^{5}$ manifold,
\begin{equation}
H =  - \sum\limits_{\left\langle {i,j} \right\rangle ,m,\sigma } ({t_{ij}^m c_{im\sigma }^\dag c_{jm\sigma }}+h.c.)+H^{(int)},
\label{eq1}
\end{equation}
where $c_{im\sigma }^\dag$ creates a hole at site $i$, orbital $m$ with spin $\sigma$. The interaction
term $H^{(int)}$ consists of the intraorbital Hubbard repulsion
$U\sum\limits_{i,m} {{n_{im \uparrow }}{n_{im
\downarrow }}}$, the interorbital interaction for opposite and parallel
spin $U'\sum\limits_{i,m \ne m'} {{n_{im \uparrow }}{n_{im' \downarrow }}}$,
$(U'-J)\sum\limits_{i,m > m',\sigma } {{n_{im\sigma
}}{n_{im'\sigma }}}$, and the Hund's
coupling $J\sum\limits_{i,m \ne m'} {\left( {c_{im \uparrow
}^\dag c_{im' \downarrow }^\dag {c_{im \downarrow }}{c_{im' \uparrow
}} + c_{im \uparrow }^\dag c_{im \downarrow }^\dag {c_{im' \downarrow
}}{c_{im' \uparrow }}} \right)}$. Because of the orbital symmetry, a well know relation $U=U'+2J$ holds.


The low-energy effective Hamiltonian of the three-orbital Hubbard model Eq.~(\ref{eq1}) is derived from the second-order perturbation processes with respect to the hopping terms in the large Hubbard interaction limit~\cite{Khaliullin2005},
\begin{eqnarray}
H_{eff} = &&\sum\limits_{\left\langle {i,j} \right\rangle }{{2} \over {{U_1}}}\left( {{S_i} \cdot {S_j} + {3 \over 4}} \right)\left[ {{A_{ij}} - {1 \over 2}\left( {{N_i} + {N_j}} \right)} \right]
\nonumber
\\
     && + {{2} \over {{U_2}}}\left( {{S_i} \cdot {S_j} - {1 \over 4}} \right)\left[ {{A_{ij}} + {1 \over 2}\left( {{N_i} + {N_j}} \right)} \right]
     \nonumber
     \\
     && + \left( {{{2} \over {{U_3}}} - {{2} \over {{U_2}}}} \right)\left( {{S_i} \cdot {S_j} - {1 \over 4}} \right){{M-1} \over M}{B_{ij}},
     \label{eq2}
\end{eqnarray}
where $U_1=U-3J, U_2=U-J, U_3=U+(M-1)J$ and $M$ is the number of orbitals, the $s=1/2$ operator $S_i$ acts on the real spin space, and the operators $A_{ij}, B_{ij}, N_i$ act on the
orbital space(It also holds for multi-orbital Hubbard model. See Supplementary Material~\cite{app}). Eq.~(\ref{eq2}) preserves the spin $SU(2)$ symmetry. Including the SOC at each site: $H_0=\lambda\sum_{i} l_{i}\cdot S_{i}$, it will split the sixfold degenerate $t_{2g}$ manifold into a $J_{eff}=1/2$ Kramers doublet and a fourfold degenerate $J_{eff}=3/2$ bands. When one hole resides at each lattice site, the physical relevant states are the half-filled Kramers doublet (isospin) with the wave function $ {1 \over {\sqrt 3 }}\left( { \left| {xy, \uparrow \downarrow } \right\rangle\pm \left| {yz, \downarrow \uparrow } \right\rangle  - i\left| {zx, \downarrow \uparrow } \right\rangle   } \right)$. Therefore, the SOC entangles the spin and orbital degrees of freedom. Consequently, the anisotropic isospin couplings may be easily realized when the orbital symmetry is broken, such as the Kitaev spin model as shown before~~\cite{Jackeli2009}.

To proceed, we introduce $U(1)$ gauge fields $(\theta_i^{yz},\theta_i^{zx},\theta_i^{xy})$ to these isospin states:
\begin{eqnarray}
 \left| { + } \right\rangle_i  &=& {1 \over {\sqrt 3 }}\left( { {e^{ + i{\theta_i^{xy}}}}\left| {xy, \uparrow } \right\rangle+ {e^{ + i{\theta _i^{yz}}}}\left| {yz, \downarrow } \right\rangle  - i{e^{ + i{\theta _i^{zx}}}}\left| {zx, \downarrow } \right\rangle   } \right) \nonumber
 \\
  \left| { - } \right\rangle_i  &=& {1 \over {\sqrt 3 }}\left( {  {e^{ - i{\theta _i^{xy}}}}\left| {xy, \downarrow } \right\rangle- {e^{ - i{\theta _i^{yz}}}}\left| {yz, \uparrow } \right\rangle  - i{e^{ - i{\theta _i^{zx}}}}\left| {zx, \uparrow } \right\rangle   } \right) \nonumber
  \\
 \label{eq3}
\end{eqnarray}
To break the orbital symmetry, we consider the situation of only two orbitals are active, for example ${t^{xy}} = t, {t^{yz}} = {t^a}t, {t^{zx}} = 0$. After projecting Eq.~(\ref{eq2}) on the isospin states Eq.~(\ref{eq3})~\cite{Jackeli2009}(see Supplementary Material~\cite{app} for details), the resulting Hamiltonian is given by $H = H_{XY}+H_H$ with,
\begin{eqnarray}
  H_{XY} &=& \sum\limits_{ < i,j > } {\left[ {{K_1}\sigma _i^x\sigma _j^x+{K_2}\sigma _i^y\sigma _j^y}+{{J_1}\sigma _i^x\sigma _j^y + {J_2}\sigma _i^y\sigma _j^x} \right]},\nonumber\\
  \label{eq4}
  \\
  H_H &=& \sum\limits_{ < i,j > }{J_{H} \widetilde{S}_i\cdot \widetilde{S}_j}.
\end{eqnarray}
$H_H$ is the Heisenberg term of the isospin: $\widetilde{S}=\vec{\sigma}/2$, and
\begin{eqnarray*}
   J_{H}=&&{\left( {{{{2t^2}} \over {9{(U-3J)}}} + {1 \over 3}{{{2t^2}} \over {9{(U-J)}}} + {2 \over 3}{{{2t^2}} \over {9{(U+2J)}}}} \right)}
   \\
   &&\times{\left( {1 + (t^a)^2 + 2t^a\cos \left( \left( \theta_i^x  - \theta_j^x \right) - \left(\theta_i^z  - \theta_j^z\right) \right)} \right)}\\
   &&+\left( {{{{2t^2}} \over {9{(U-3J)}}} - {{{2t^2}} \over {9{(U-J)}}}} \right)(1 - (t^a)^2).
\end{eqnarray*}
Under the condition: i) $t^a=1$ and  $\left( \theta_i^{yz}  - \theta_j^{yz} \right) - \left(\theta_i^{xy}  - \theta_j^{xy}\right)=\pi$; or ii) $t^a=-1$ and  $\left( \theta_i^{yz}-\theta_j^{yz} \right) - \left(\theta_i^{xy}-\theta_j^{xy}\right)=0$, the Heisenberg term can be eliminated. In the case of two-site periodical gauge fields, the coefficients are specified as $K_1=X-A, K_2=X+A, J_1=B-(-1)^iY, J_2=B+(-1)^iY$~\cite{app}.

Next let us study the topological properties of the effective isospin model $H_{XY}$. This model differs from the usual $XY$ chain in that it consists of the exchange couplings between the $x$ and $y$ spin components.
It can be solved exactly by mapping the isospin operators to Majorana fermions
using the Jordan-Wigner transformation~\cite{Feng2007,Sen1997},
\begin{equation}
\sigma _i^x = (\prod\limits_{j < i} {{\rm{i}}{a_j}{b_j}}) {a_i},
\sigma _i^y = (\prod\limits_{j < i} {{\rm{i}}{a_j}{b_j}}) {b_i},
\sigma _i^z = {\rm{i}}{b_i}{a_i},
\end{equation}
where ${a_i}$,${b_i}$ are Majorana operators on the $i$ site. Now the Hamiltonian can be rewritten as
\begin{equation}
H_{XY} = {\rm{i}}\sum\limits_i {\left[ {J_2{a_i}{a_{i + 1}} - J_1{b_i}{b_{i + 1}}}{-K_1{b_i}{a_{i + 1}} + K_2{a_i}{b_{i + 1}}} \right]},
\label{eq9}
\end{equation}
Compared to the 1D spinless $p$-wave superconductor system~\cite{Kitaev2001,Alicea2011,DeGottardi2011}, the Hamiltonian $H_{XY}$ [Eq.~(\ref{eq9})] has an inter-site coupling of the same species of Majoranas, but it has no intra-site coupling. We will argue that these differences are essential to realize the 1D braiding of two Majorana fermions in the followings.

\begin{figure}
  \centering
  \includegraphics[width=0.3\textwidth]{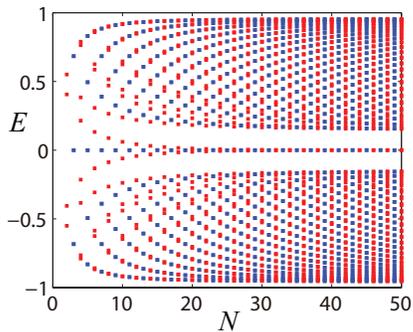}
  \caption{\label{effect}The energy levels calculated using Eq.~(\ref{eq9}) {\it v.s.} the number $N$ of the lattice sites. The blues denote the chain with an odd number of sites and the reds an even number of the lattice sites.}
\end{figure}

$H_{XY}$ preserves both the particle-hole symmetry and an anti-unitary symmetry $\sigma^x\mathcal{K}$ with $(\sigma^x\mathcal{K})^2=1$($\sigma^x=\prod\sigma_i^x$ and $\mathcal{K}$ is the complex conjugation, so we can call it pseudo time-reversal symmetry). According to the general topological classification~\cite{Tewari2012,Ryu2010}, the system belongs to the class BDI characterized by a $Z$ invariant.
It is believed that there are topological protected bound states on topological defects, i.e. domain walls between different topological regions in 1D, and if they have zero energy, they are Majorana zero modes. To test the presence of the zero-energy modes, we turn to the transfer matrix method which is specially suitable for a 1D system~\cite{Hatsugai1993,DeGottardi2011}. We can get a transfer equation with respect to the energy and the wave function of an excitation mode~\cite{app}
$\left( {\matrix{
   {{\varphi _{i+1}}}  \cr
   {{\varphi _i}}  \cr
 } } \right) = {T_i\left( {\varepsilon} \right)}\left( {\matrix{
   {{\varphi _i}}  \cr
   {{\varphi _{i - 1}}}  \cr
 } } \right),$ where $\varphi _i=(a_{i},b_{i})^{T}$.
With an open boundary condition ${\varphi _0} ={\varphi _{N+ 1}} = 0$, the necessary condition to get a physical solution is that the determinant of coefficient of transfer matrix should be zero,
\begin{equation}
 {\textrm{Det}}\left[ {{{\left[ {\prod\limits_{i = 1}^{N} {{T_i\left( {\varepsilon} \right)}} } \right]}_{11}}} \right] = 0.
 \label{eq11}
\end{equation}
 ($ \left[  M\right]_{11}$ means the upper left $2\times2$ matrix.)
The system having Majorana zero modes requires that Eq.~(\ref{eq11}) holds when $\varepsilon=0$. A straightforward calculation shows that
\begin{equation}
 Q=\textrm{Det}\left[ {{{\left[ {\prod\limits_{i = 1}^N {{T_i}\left( {\varepsilon  = 0} \right)} } \right]}_{11}}} \right] = \left\{ \matrix{
  0,N =\rm odd \hfill \cr
  1,N =\rm even \hfill \cr}  \right.
\end{equation}
Therefore the existence or the absence of a gap between the bound states is determined by the quantity $Q$ whose module is gauge invariant. So we can define $Q$ as a $Z_2$ invariant when the system is topological nontrivial in the class BDI.
It is that $Q=0$, which is nontrivial, ensures the presence of the two Majorana zero modes, if and only if the chain has odd number of lattice sites.
In the other case of $Q=1$, i.e. one has even number of lattice sites, two Majorana zero modes will couple and open a gap, so they are no longer Majorana zero modes.
We note that this topological classification is also applied to a class of the particle-hole symmetrical 1D or quasi-1D topological system characterized by a $Z$ invariant. For example, in Haldane model~\cite{Haldane1988} the two edge states on the opposite sides of a ribbon hexagon lattice will be topological protected from opening a gap as long as the ribbon has odd number of layers, no matter how close the two edges become. The above analytical analysis can be demonstrated numerically as shown in Fig.~\ref{effect}. It shows that the energy of the bound edge(end) states always remain zero for an odd number of lattice sites in the chain. However, in the case of even number of lattice sites the bound edge states will couple and a noticeable gap emerges when the length of the chain is decreased, though the gap approaches to zero in the infinite limit. In the yet proposed schemes to realize the Majorana zero modes, such as the 1D nanowires with proximity coupling to an $s$-wave SC~\cite{Sau2010,Oreg2010}, the size of the nanowires is limited in the microfabrication. Therefore, this property poses a severe limit on the realization of exact Majorana fermions in the nanowires.

\begin{figure}
  \centering
  \includegraphics[width=0.47\textwidth]{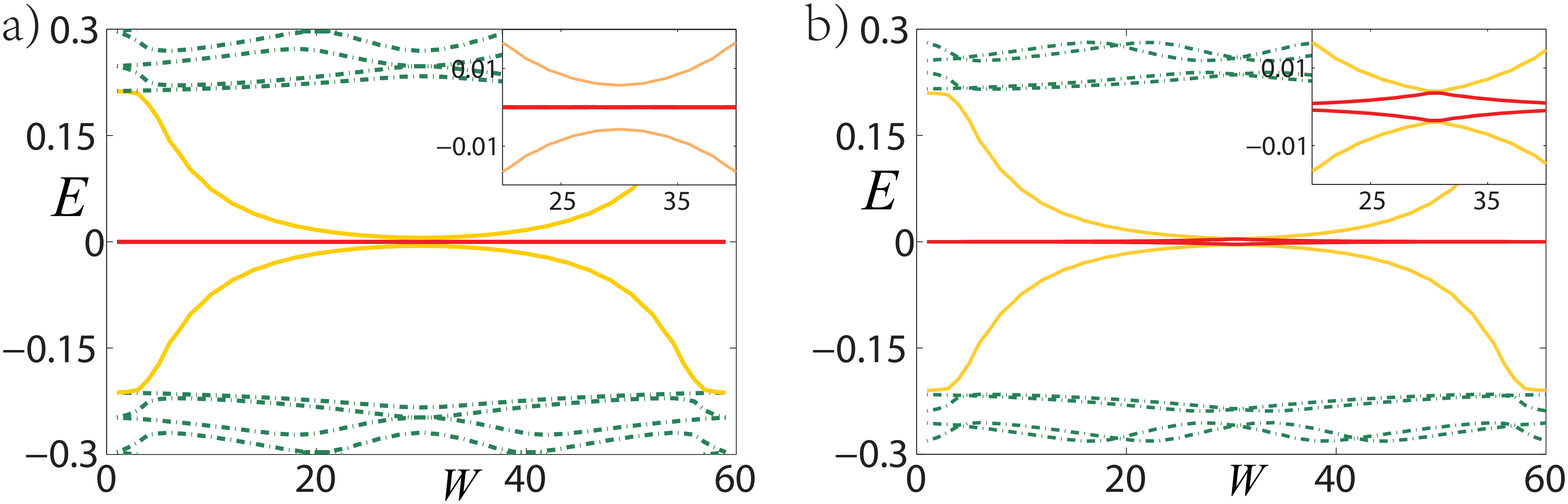}
  \caption{\label{braiding}The energy level as a function of the position of the domain wall $W$ in the 1D chain. a) is for 59 and b) for 60 lattice sites. The greens represent the bulk states, the reds represent two Majorana fermions, and the oranges represent the states resulting from the coupling of the other two bound states. Inset: the enlargement around the zero energy.}
\end{figure}

\emph{Braiding}---Now let us check if it is possible to exchange the two Majoranas which obey the braiding statistics in this strictly 1D geometry. To exchange the two Majoranas, we will introduce a domain wall to bound Majoranas as usual and the movement of the domain wall will carry the Majoranas. In our case, a domain wall can be created by switching the active orbital from $yz$ to $zx$ i.e. ${t^{yz}} = 0, {t^{zx}} = {t^a}t$. It corresponds to a shift of the parameters $A$ and $B$ to $-A$ and $-B$, which results in the winding number of system from $\textbf{1}$ to $\textbf{-1}$~\cite{Qi2008}. Consequently, four bound states will emerge in the system, two at the domain wall between $\textbf{1}$ and $\textbf{-1}$ and two at both ends of the chain. However, all four bound states are not presumed to be Majorana zero modes because of the coupling between them in the finite system. Indeed, as shown above, at least a pair of Majoranas is protected from fusing by the new topological invariant $Q=0$ as long as the chain has odd number of sites.
This property can further be demonstrated numerically in Fig.~\ref{braiding}, where the energy level as a function of the position of the domain wall $W$ is presented. One can find that in fact there are only two of the four bound states are Majoranas in the case of odd number of sites (Fig.~\ref{braiding} a)). When the domain wall moves, the energy of the other two bound states approaches to zero gradually, but they will never intersect with the two Majorana zero modes. On the other hand, in the case of even number of sites, even the chain is long enough so that the two bound state might be approximated as Majorana zero modes, the energy levels of the other two bound states will eventually intersect with these approximate zero modes and it opens a gap as shown in Fig.~\ref{braiding} b).

\begin{figure}
  \centering
  \includegraphics[width=0.47\textwidth]{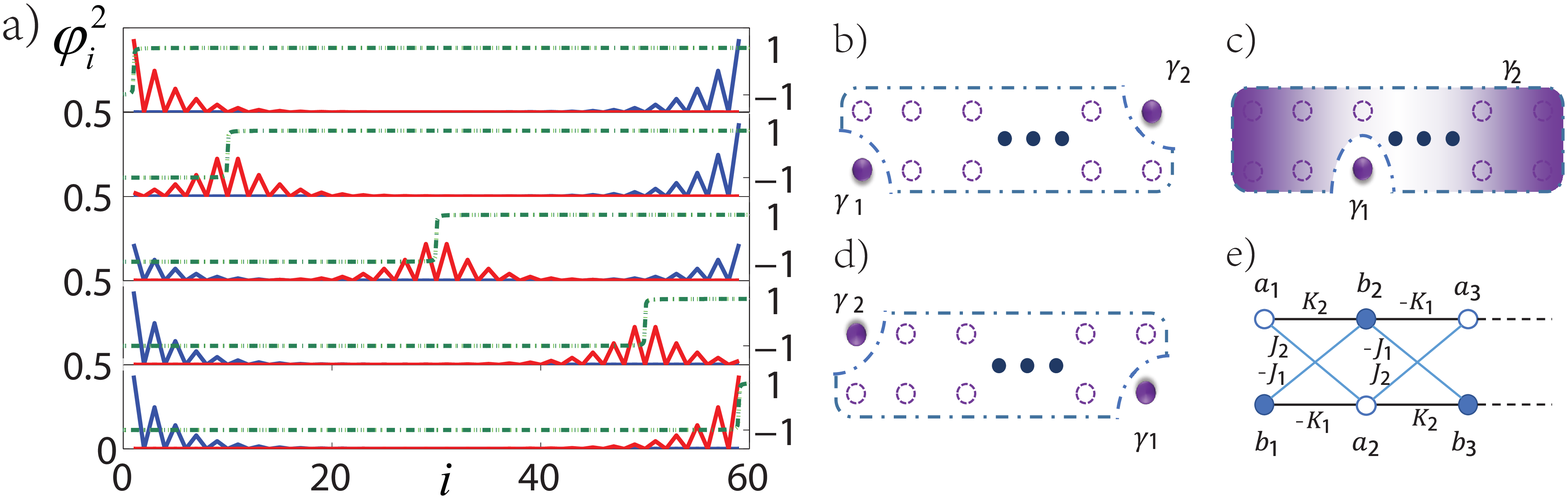}
  \caption{\label{braiding1}a) The distribution of the wavefunctions of two Majorana fermions(the red and the blue lines) on the 1D spin chain with 59 lattice sites. The green dashed line indicates the winding number, so the step indicates the position of the domain wall. b)-d)illustrate the physical picture of the exchanging process of two Majoranas $\gamma_1,\gamma_2$ in the geometrical 1D spin chain. e)Pictorial representation of the couplings between Majorana fermions as described by the Hamiltonian Eq.~(\ref{eq9}).}
\end{figure}

Now let us figure out how does this pair of Majoranas evolve in the chain with an odd number of sites. In the adiabatic approximation, the system remains in the instantaneous eigenstate of the Hamiltonian, and we can obtain the pair of Majoranas as,
\begin{equation}
  {  \gamma _1}(W) = \sum\limits_{0 \le i < W/2} {{P^i}{{  \mu }_{2i + 1}}}  + \sum\limits_{W/2 \le i < N/2} {{P^{W - i - 1}}{{  \mu }_{2i + 1}}},
   \label{eq12}
\end{equation}
\begin{equation}
  {  \gamma _2}(W) = \sum\limits_{0 \le i < W/2} {{P^{ - i}}{{  \upsilon }_{2i + 1}}}  + \sum\limits_{W/2 \le i < N/2} {{P^{ - W + i + 1}}{{  \upsilon }_{2i + 1}}},
   \label{eq13}
\end{equation}
where $P>1$, $\mu_i$ and $\upsilon_i$ are the linear combinations of the Majorana operators $a_i$ and $b_i$(see Supplementary Material~\cite{app}). One can find that ${\gamma _1}$ assembles around the domain wall at the position $W$, while ${\gamma _2}$ distributes around the two ends. When the domain wall transports, the evolution of the two Majoranas is exactly described by the formula Eq.~(\ref{eq12}) and (\ref{eq13}). To show clearly this process, we present the numerical results of Eq.~(\ref{eq12}) and (\ref{eq13}) in Fig.~\ref{braiding1} a). As shown, when the domain wall is near the left end two Majoranas distribute around the two ends, respectively. With the transport of the domain wall, the Majorana $\gamma_1$ carried by the domain wall moves along the chain. At the meantime, the spectral weight of the Majorana $\gamma_2$ transfers gradually to the left end. When $\gamma_1$ arrives at the right end, $\gamma_2$ completely transfers to the left end. However, to complete the braiding, we finally need to do a gauge transformation which transforms the Kramers doublet $\left|  \pm  \right\rangle $ into $\exp ( \pm i{\pi  \over 4})\left|  \pm  \right\rangle $. It results in,
\begin{eqnarray}
  {\gamma _1}'\left( N \right) &=& \sum\limits_{0 \le i < N/2} {{P^i}{{ \upsilon }_{2i + 1}}}  =  - {\gamma _2}\left( 1 \right), \nonumber\\
  {\gamma _2}'\left( N \right) &=& {\gamma _1}\left( 1 \right),
\end{eqnarray}
where $'$ denotes the states after the gauge transformation. Thus, we realize the braiding of two Majoranas: ${\gamma _1} \to  - {\gamma _2},{\gamma _2} \to {\gamma _1}$. In this process, the two Majoranas avoid a catastrophic encounter magically, so we realize the braiding in the strictly 1D geometry, i.e., in a spin chain. This surprised result can be understood if we look at the process in the channel of Majoranas, as shown pictorially in Fig.3(b-d). Because there are two species of Majoranas $a_{i},b_{i}$, in fact what the Majoranas ${\gamma _1}, {\gamma _2}$ (a linear combination of $a_{i},b_{i}$) see is two parallel chains.
We speculate that the absence of the coupling between different species $a_{i},b_{i}$ on the same sites and the presence of the coupling of the same species between the nearest-neighbors $a_{i},a_{i+1}$ or $b_{i},b_{i+1}$ in our model (see Fig.~\ref{braiding1} e)) might be essential for the braiding in this geometric 1D spin chain.

In our solution, we find one exception that, when ${A^2} + {B^2} = {X^2} + {Y^2}$, the four bound states are all Majorana fermions and the braiding will break down(see the Supplementary  materials for detail). We notice that a simple protocol that permits adiabatic exchange of two Majorana fermions in 1D superconductor wire has been proposed~\cite{Chiu2014}. The exchange of two Majorana fermions relies on a domain wall in the superconducting order parameter which hosts a pair of ancillary Majoranas.



\emph{Speculation and discussion}---To get the exactly solvable 1D spin model $H_{XY}$, we have introduced the $U(1)$ gauge fields $(\theta_{i}^{yz},\theta_{i}^{zx},\theta_{i}^{xy})$ to Kramers doublet. These gauge fields can be implemented by a rotation operation $R\left( {\alpha ,\beta } \right) = \exp \left( {{\rm{i}}{{\hat L}_z}\alpha  + 2{\rm{i}}{{\hat J}_z}\beta } \right)$, where ${{{\hat L}_z}}$ is the $z$ component of the effective orbital angular momentum and ${{{\hat J}_z}}$ is the $z$ component of the total angular momentum of the spin and orbital. In this way, we have $(\theta_{i}^{yz},\theta_{i}^{zx},\theta_{i}^{xy})=(\alpha-\beta/2,\alpha-\beta/2,\beta/2)$. $R\left( {\alpha ,0} \right)$ is the rotation operator acting on the orbital space, which is supposed to rotate the crystal field inversely. It could be realized by rotating the oxygen octahedron of transition metal compounds along the $z$-axis in solids. On the other hand, $R\left( {0 ,\beta} \right)$, a rotation operator acting on the whole space, is difficult to realize because we are not able to rotate a spin. We note that this operation can be replaced by employing an external magnetic field which amounts to introduce a dynamical phase into $U\left|  \pm  \right\rangle $, with $U=\exp \left( {{-\rm{i}}\int_0^T {{\pm E}\left( t \right)dt} } \right)$~\cite{app}. Admittedly, the rotation operations and the implementation to create the domain wall by shifting the active orbital are in fact difficult to realize in solids. However, thanks to the recent rapid developments on the quantum simulation by using a well-controlled quantum system to simulate complex quantum matter, we propose to realize this scheme either in cold atom systems~\cite{Duan2003} or in quantum simulators consists of superconducting circuits~\cite{Houck2012}.

Finally, we would like to discuss the possible effect of the Heisenberg term if it has not been eliminated. In this case, an additional term which will enter the Hamiltonian $H_{XY}$ is $\sigma_i^z\sigma_{i+1}^z$. It is easy to see that the ground state is still Kramers doubly degenerate if the lattice site $N$ is odd because of its time-reversal symmetry $(\prod \textrm{i}\sigma_i^y\mathcal{K})^2=-1$. According to Ref.~\cite{Wei2014}, the ground state of a class of interacting Majorana fermion models is always doubly degenerate if $N$ is odd which is related by a particle-hole operation. Thus, it is interesting to generalize this conclusion to our system when the term $\sigma_i^z\sigma_{i+1}^z$ is added, and to investigate if its Kramers doubly degeneracy is related to a pair of Majorana zero modes.


\emph{Summary}---In conclusion, we derive an exactly solvable one-dimensional spin model from the three-orbital Hubbard model with a strong spin-orbit coupling by projecting it onto the isospin states with $U(1)$ gauge fields. It has a topological nontrivial phase characterized by Majorana end modes which are protected by a new $Z_2$ topological invariant related to the parity of the lattice sites (odd or even number of sites).  With the protection of this new topological invariant $Q = 0$, we realize the braiding of two Majoranas in this strictly geometric one-dimensional spin chain.

\begin{acknowledgments}
This work was supported by the National Natural Science Foundation of China (91021001, 11190023 and 11204125) and the Ministry of Science and Technology of China (973 Project Grants No.2011CB922101 and No. 2011CB605902).
\end{acknowledgments}


\newpage
\widetext
\vspace{0.5cm}
\appendix
\begin{center}
\textbf{\large Supplementary  materials}
\end{center}

\section{Appendix A: From Hubbard model to the effective spin model}\label{A}

The multi-orbital Hubbard model can be expressed as,
\begin{eqnarray*}
  H &=& H_0+H_1, \\
  H_1 &=& - \sum\limits_{\left\langle {i,j} \right\rangle ,m,\sigma } ({t_{ij}^m c_{im\sigma }^\dag c_{jm\sigma }}+h.c.), \\
  H_0 &=&U\sum\limits_{i,m} {{n_{i,m \uparrow }}{n_{i,m \downarrow }}}  + U'\sum\limits_{i,m \ne m'} {{n_{i,m \uparrow }}{n_{i,m' \downarrow }}}  + (U' - J)\sum\limits_{i,m > m',\sigma } {{n_{i,m\sigma }}{n_{i,m'\sigma }}}\\
  &&+J\sum\limits_{i,m \ne m'} {\left( {c_{i,m \uparrow }^\dag c_{i,m' \downarrow }^\dag {c_{i,m \downarrow }}{c_{i,m' \uparrow }} + c_{i,m \uparrow }^\dag c_{i,m \downarrow }^\dag {c_{i,m' \downarrow }}{c_{i,m' \uparrow }}} \right)}.
\end{eqnarray*}
In the limit of strong Hubbard correlation $U \gg t$, it is well known that we can derive the Heisenberg model from the one-band Hubbard model when the system is at half-filling. This is carried out by the second-order perturbation processes with respect to the transfer term, which can be applied to the above multi-orbital model. Thus, treating $H_1$ as a perturbation, we obtain an effective Hamiltonian which could be expanded in Taylor series,
\begin{equation}
  H_{mm'}^{eff} = \left\langle m \right|{H_0}\left| {m'} \right\rangle  + \left\langle m \right|{H_1}\left| {m'} \right\rangle  + {1 \over 2}\sum\limits_l {\left( {{{\left\langle m \right|{H_1}\left| l \right\rangle \left\langle l \right|{H_1}\left| {m'} \right\rangle } \over {{E_m} - {E_l}}} + {{\left\langle m \right|{H_1}\left| l \right\rangle \left\langle l \right|{H_1}\left| {m'} \right\rangle } \over {{E_{m'}} - {E_l}}}} \right)}  +  \cdots,
\end{equation}
where $\left| n \right\rangle $ and $E_n$ are the eigenstates and eigenvalue of $H_0$. The first term is the unperturbed Hamiltonian $H_0$, the second term is the first order correction and the third term is the second order correction.

Because of a large $U$, the Hilbert space of $H_0$ is separated into the zero-energy subspace containing states with empty or singly occupied sites, and the other high-energy subspace with multi-particle occupied sites. We consider the case that there is only one electron(hole) per site, so that the first order correction vanishes in the low-energy approximation. Since there is a large gap between low-energy excitations and high-energy excitations in large $U$ limit, we can discuss the effective Hamiltonian in the zero-energy subspace. In this way, the effective Hamiltonian can be written as,
\begin{equation}
H_{ss'}^{eff} =  - \sum\limits_{\left\langle {i,j} \right\rangle,m,\sigma } {\sum\limits_{\left\langle {i,j} \right\rangle',m',\sigma'} {\sum\limits_d {{{\left\langle s \right|\left( {t_{ij}^mc_{im\sigma }^\dag {c_{jm\sigma }} + h.c.} \right)\left| d \right\rangle \left\langle d \right|\left( {t_{ij}^mc_{im\sigma }^\dag {c_{jm\sigma }} + h.c.} \right)'\left| {s'} \right\rangle } \over {{E_d}}}} } },
\label{eqS1}
\end{equation}
where ${\left| {s} \right\rangle }$ and ${\left| {s'} \right\rangle }$ denotes single-occupied states.  
Since ${\left| d \right\rangle }$ represents a state with only a doubly occupied site, we can get ${{E_d}}$ by simply calculating an onsite $H_0$ in two particles space:
\begin{eqnarray*}
  E_d &:& {\left| d \right\rangle_i }\\
  U'-J &:& \left| {m \uparrow m' \uparrow } \right\rangle ,\left| {m \downarrow m' \downarrow } \right\rangle ,\left| {m \uparrow m' \downarrow } \right\rangle  + \left| {m' \downarrow m \uparrow } \right\rangle \\
  U'+J &:& \left| {m \uparrow m' \downarrow } \right\rangle  - \left| {m' \downarrow m \uparrow } \right\rangle \\
  U-J &:&\left| {m \uparrow m \downarrow } \right\rangle  - \left| {m' \downarrow m' \uparrow } \right\rangle \\
  U+(M-1)J &:& \sum\limits_m {\left| {m \uparrow m \downarrow } \right\rangle }
\end{eqnarray*}
When the orbitals preserve $SO(3)$ symmetry, there is $U'=U-2J$. After a detail calculation, we can write the effective Hamiltonian in the second-quantization representation:
\begin{eqnarray}
H_{eff} = &&\sum\limits_{\left\langle {i,j} \right\rangle }{{2} \over {{U_1}}}\left( {{S_i} \cdot {S_j} + {3 \over 4}} \right)\left( {{A_{ij}} - {1 \over 2}\left( {{N_i} + {N_j}} \right)} \right)     \nonumber\\
     && + {{2} \over {{U_2}}}\left( {{S_i} \cdot {S_j} - {1 \over 4}} \right)\left( {{A_{ij}} + {1 \over 2}\left( {{N_i} + {N_j}} \right)} \right) \nonumber \\
     && + \left( {{{2} \over {{U_3}}} - {{2} \over {{U_2}}}} \right)\left( {{S_i} \cdot {S_j} - {1 \over 4}} \right){{M-1} \over M}{B_{ij}},
     \label{eqS2}
\end{eqnarray}
where, $M$ is the number of orbitals and $U_1=U-3J, U_2=U-J, U_3=U+(M-1)J$. The spin-1/2 operator $S_i$ acts on the real spin space, and $A_{ij}, B_{ij}, N_i$ act on the orbital space,
\begin{eqnarray*}
  {A_{ij}} &=& \sum\limits_{m,m'} {t_{ij}^mt_{ji}^{m'}c_{im}^\dag {c_{im'}}c_{jm'}^\dag {c_{jm}}},\\
  {B_{ij}} &=& \sum\limits_{m,m'} {t_{ij}^mt_{ji}^{m'}c_{im}^\dag {c_{im'}}c_{jm}^\dag {c_{jm'}}},\\
  {N_i} &=& \sum\limits_m {t_{ij}^mt_{ji}^mc_{im}^\dag {c_{im}}},\\
  c_{im\sigma }^\dag {c_{im'\sigma '}} &=& c_{im}^\dag {c_{im'}}\left( {{1 \over 2}{\delta _{\sigma \sigma '}} + {S_i} \cdot {{\vec \sigma }_{\sigma \sigma '}}} \right).
\end{eqnarray*}

With a strong spin-orbital coupling(SOC), a part of singly-occupied states is lifted from zero energy. So, to get the low-energy effective Hamiltonian, we need to project Eq.~(\ref{eqS2}) on the lowest levels of the SOC Hamiltonian~\cite{Jackeli2009}. In the paper, we consider the $t_{2g}^{5}$ manifold of $xy,xz,yz$ orbitals for which the low-energy electronic properties can be described in terms of $s=1/2$ isospin states $\left| \pm  \right\rangle$. We further assume that only two orbitals $x,z$ are active(hereafter we use the abbreviation $yz\rightarrow x,zx\rightarrow y,xy\rightarrow z$), i.e., ${t^{z}} = t, {t^{x}} = {t^a}t, {t^{y}} = 0$. After introducing the $U(1)$ gauge fields to $\left| \pm  \right\rangle$, we project Eq.(\ref{eqS2}) on the Kramers doublet with the operator $P_{ij}=( {{{\left|  +  \right\rangle }_i} + {{\left|  -  \right\rangle }_i}} )( {{{\left|  +  \right\rangle }_j} + {{\left|  -  \right\rangle }_j}} )( {{{\left\langle  +  \right|}_j} + {{\left\langle  -  \right|}_j}} )( {{{\left\langle  +  \right|}_i} + {{\left\langle  -  \right|}_i}} )$,
\begin{eqnarray*}
  {P_{ij}}\left( {{S_i} \cdot {S_j} + {3 \over 4}} \right){A_{ij}}{P_{ij}} &=& {{{t^2}} \over {36}}[ {4\left( {3 + 3{{({t^a})}^2} + 2{t^a}\cos \left( {{\Delta ^{xx}} - {\Delta ^{zz}}} \right)} \right) + \left( {1 + {{({t^a})}^2} + 2{t^a}\cos \left( {{\Delta ^{xx}} - {\Delta ^{zz}}} \right)} \right)\sigma _i^z\sigma _j^z} \\
   &&+ \left( {\cos \left( {2{\Delta ^{zz}}} \right) + {{\left( {{t^a}} \right)}^2}\cos \left( {2{\Delta ^{xx}}} \right) + 4{t^a}\cos \left( {{\Delta ^{xx}} + {\Delta ^{zz}}} \right)} \right)\left( {\sigma _i^y\sigma _j^y + \sigma _i^x\sigma _j^x} \right)\\
    &&+ \left( {2{t^a}\cos \left( {{\Sigma ^{xx}} + {\Sigma ^{zz}}} \right)} \right)\left( {\sigma _i^y\sigma _j^y - \sigma _i^x\sigma _j^x} \right)\\
   &&- \left( {\sin \left( {2{\Delta ^{zz}}} \right) + {{\left( {{t^a}} \right)}^2}\sin \left( {2{\Delta ^{xx}}} \right) + 4{t^a}\sin \left( {{\Delta ^{xx}} + {\Delta ^{zz}}} \right)} \right)\left( {\sigma _i^y\sigma _j^x - \sigma _i^x\sigma _j^y} \right)\\
    &&- \left( {2{t^a}\sin \left( {{\Sigma ^{xx}} + {\Sigma ^{zz}}} \right)} \right)\left( {\sigma _i^y\sigma _j^x + \sigma _i^x\sigma _j^y} \right)]\\
  {P_{ij}}\left( {{S_i} \cdot {S_j} - {1 \over 4}} \right){A_{ij}}{P_{ij}} &=& {{{t^2}} \over {36}}[ {4\left( { - 1 - {{({t^a})}^2} + 2{t^a}\cos \left( {{\Delta ^{xx}} - {\Delta ^{zz}}} \right)} \right) + \left( {1 + {{({t^a})}^2} + 2{t^a}\cos \left( {{\Delta ^{xx}} - {\Delta ^{zz}}} \right)} \right)\sigma _i^z\sigma _j^z} \\
   &&+ \left( {\cos \left( {2{\Delta ^{zz}}} \right) + {{\left( {{t^a}} \right)}^2}\cos \left( {2{\Delta ^{xx}}} \right)} \right)\left( {\sigma _i^y\sigma _j^y + \sigma _i^x\sigma _j^x} \right) - \left( {2{t^a}\cos \left( {{\Sigma ^{xx}} + {\Sigma ^{zz}}} \right)} \right)\left( {\sigma _i^y\sigma _j^y - \sigma _i^x\sigma _j^x} \right)\\
   &&- \left( {\sin \left( {2{\Delta ^{zz}}} \right) + {{\left( {{t^a}} \right)}^2}\sin \left( {2{\Delta ^{xx}}} \right)} \right)\left( {\sigma _i^y\sigma _j^x - \sigma _i^x\sigma _j^y} \right) + \left( {2{t^a}\sin \left( {{\Sigma ^{xx}} + {\Sigma ^{zz}}} \right)} \right)\left( {\sigma _i^y\sigma _j^x + \sigma _i^x\sigma _j^y} \right)]
\end{eqnarray*}
\begin{eqnarray*}
  {P_{ij}}\left( {{S_i} \cdot {S_j} - {1 \over 4}} \right){B_{ij}}{P_{ij}} &=& {{{t^2}} \over {36}}[ {4\left( { - 1 - {{({t^a})}^2} - 2{t^a}\cos \left( {{\Delta ^{xx}} - {\Delta ^{zz}}} \right)} \right) + \left( {1 + {{({t^a})}^2} + 2{t^a}\cos \left( {{\Delta ^{xx}} - {\Delta ^{zz}}} \right)} \right)\sigma _i^z\sigma _j^z} \\
   &&+ \left( {\cos \left( {2{\Delta ^{zz}}} \right) + {{\left( {{t^a}} \right)}^2}\cos \left( {2{\Delta ^{xx}}} \right) + 2{t^a}\cos \left( {{\Delta ^{xx}} + {\Delta ^{zz}}} \right)} \right)\left( {\sigma _i^y\sigma _j^y + \sigma _i^x\sigma _j^x} \right)\\
   &&- \left( {\sin \left( {2{\Delta ^{zz}}} \right) + {{\left( {{t^a}} \right)}^2}\sin \left( {2{\Delta ^{xx}}} \right) + 2{t^a}\sin \left( {{\Delta ^{xx}} + {\Delta ^{zz}}} \right)} \right)\left( {\sigma _i^y\sigma _j^x - \sigma _i^x\sigma _j^y} \right)] \\
  {P_{ij}}\left({S_i} \cdot {S_j}\right){(N_i+N{j})/2}{P_{ij}} &=&{{{{ {t}}^2}} \over {72}}[ 2\left( (t^a)^2-1 \right)\sigma _i^z\sigma _j^z\\
  &&+\left( {2(t^a)^2\cos \left( {2{\Delta ^{xx}}} \right) + 2\cos \left( {2{\Delta ^{zz}}} \right) - (t^a)^2\cos \left( {2{\Delta ^{xy}}} \right) - (t^a)^2\cos \left( {2{\Delta ^{yx}}} \right)} \right)\\
  &&\times\left( {\sigma _i^y\sigma _j^y + \sigma _i^x\sigma _j^x} \right)\\
     &&+ \left( {(1+(t^a)^2)\cos \left( {2{\Sigma ^{xz}}} \right) + (1+(t^a)^2)\cos \left( {2{\Sigma ^{zx}}} \right) - \cos \left( {2{\Sigma ^{yz}}} \right) - \cos \left( {2{\Sigma ^{zy}}} \right)} \right)\\
     &&\times\left( {\sigma _i^y\sigma _j^y - \sigma _i^x\sigma _j^x} \right)\\
    &&- \left( {2(t^a)^2\sin \left( {2{\Delta ^{xx}}} \right) + 2\sin \left( {2{\Delta ^{zz}}} \right) - (t^a)^2\sin\left( {2{\Delta ^{xy}}} \right) - (t^a)^2\sin\left( {2{\Delta ^{yx}}} \right)} \right)\\
    &&\times\left( {\sigma _i^y\sigma _j^x - \sigma _i^x\sigma _j^y} \right)\\
    &&- \left( {(1+(t^a)^2)\sin \left( {2{\Sigma ^{xz}}} \right) + (1+(t^a)^2)\sin \left( {2{\Sigma ^{zx}}} \right) - \sin\left( {2{\Sigma ^{yz}}} \right) - \sin\left( {2{\Sigma ^{zy}}} \right)} \right)\\
    &&\times\left( {\sigma _i^y\sigma _j^x + \sigma _i^x\sigma _j^y} \right)] \\
  {P_{ij}}{(N_i+N{j})/2}{P_{ij}} &=& {{{(1+{\left( {{t^a}} \right)}^2)t^2}} \over {3}}
\end{eqnarray*}
with ${{\Delta ^{mm'}}}=\theta _i^m - \theta _j^{m'}$ and ${{\Sigma ^{mm'}}}=\theta _i^m + \theta _j^{m'}$. Then, we arrive at $H = H_{XY}+H_H$,
\begin{eqnarray}
  H_{XY} &=& \sum\limits_{ < i,j > } {\left[ {{K_1}\sigma _i^x\sigma _j^x+{K_2}\sigma _i^y\sigma _j^y} \right]}+\sum\limits_{ < i,j > } {\left[ {{J_1}\sigma _i^x\sigma _j^y + {J_2}\sigma _i^y\sigma _j^x} \right]},
  \nonumber
  \\
  H_H &=& \sum\limits_{ < i,j > }{J_{H< i,j >} \widetilde{S}_i\cdot \widetilde{S}_j}.
 \nonumber
\end{eqnarray}
Where $\widetilde{S}$ is the isospin $\widetilde{S}=(\sigma^x/2, \sigma^y/2, \sigma^z/2)$, and
 \begin{eqnarray*}
   J_{H< i,j >}=&&{\left( {{{{2t^2}} \over {9({U-3J})}} + {1 \over 3}{{{2t^2}} \over {9({U-J})}} + {2 \over 3}{{{2t^2}} \over {9({U+2J})}}} \right)}\cdot{\left( {1 + (t^a)^2 + 2t^a\cos \left( \left( \theta_i^x  - \theta_j^x \right) - \left(\theta_i^z  - \theta_j^z\right) \right)} \right)}\\
   &&+\left( {{{{2t^2}} \over {9{(U-3J)}}} - {{{2t^2}} \over {9{(U-J)}}}} \right)(1 - (t^a)^2)
 \end{eqnarray*}
It is tedious to write down the full forms of the parameters $K_1$, $K_2$, $J_1$, and $J_2$, which in fact can be obtained from the formulas above. Instead, we list another set of parameters
$X$, $Y$, $A$, and $B$, via $K_1=X-A, K_2=X+A, J_1=B-(-1)^iY, J_2=B+(-1)^iY$. These are the coefficients of the Hamiltonian of the system with a two-site period $U(1)$ gauge field, i.e., odd sites satisfy $(\theta_o^x,\theta_o^y,\theta_o^z)=(\alpha_o,\alpha_o,0)$, and even sites satisfy $(\theta_e^x,\theta_e^y,\theta_e^z)=(\alpha_e,\alpha_e,0)$.
\begin{eqnarray}
 X &= &\left( {{{{t^2}t^a} \over {9{U_1}}} - {{{t^2}t^a} \over {9{U_2}}}} \right)\cos \left( {\alpha_o  - \alpha_e } \right)
      -  \left( {{{{t^2}(t^a)^2} \over {9{U_1}}} + {{{t^2}(t^a)^2} \over {9{U_2}}}} \right){\sin ^2}\left( {\alpha_o  - \alpha_e } \right)
     -  { {{t^2}(2-(t^a)^2)}\over{18}}\left( {{ {1}\over {{U_1}}} - {1 \over 3}{{1} \over {{U_2}}}-{2 \over 3}{{1} \over {{U_3}}}} \right)
     \nonumber
          \\
 Y&=&-{{2t^2t^a}\over{9U_1}}\sin\left(\alpha_o-\alpha_e\right)-\left({{t^2(t^a)^2}\over{18U_1}}+{{t^2(t^a)^2}\over{18U_2}}\right)\sin\left(2(\alpha_o-\alpha_e)\right)
    \nonumber
    \\        
 A&=&{ {{t^2}t^a}\over{9}}\left( {{ {1}\over {U_1}} - {{1} \over {U_2}}} \right)\cos \left( \alpha_o+\alpha_e\right)-{ {{t^2}(t^a)^2}\over{18}}\left( {{ {1}\over {{U_1}}} - {1 \over 3}{{1} \over {{U_2}}}-{2 \over 3}{{1} \over {{U_3}}}} \right)\cos \left( \alpha_o-\alpha_e\right)\cos \left( \alpha_o+\alpha_e\right)
   \nonumber
   \\
 B&=&-{ {{t^2}t^a}\over{9}}\left( {{ {1}\over {U_1}} - {{1} \over {U_2}}} \right)\sin \left( \alpha_o+\alpha_e\right)+{ {{t^2}(t^a)^2}\over{18}}\left( {{ {1}\over {{U_1}}} - {1 \over 3}{{1} \over {{U_2}}}-{2 \over 3}{{1} \over {{U_3}}}} \right)\cos \left( \alpha_o-\alpha_e\right)\sin \left( \alpha_o+\alpha_e\right).
   \nonumber
 \end{eqnarray}

\section{Appendix B: Method of transfer matrix}\label{C}
The transfer matrix  ${L_i}\left( {\begin{array}{*{20}{c}}
  {{\varphi _{i + 1}}} \\
  {{\varphi _i}}
\end{array}} \right) = {R_i}\left( {\begin{array}{*{20}{c}}
  {{\varphi _i}} \\
  {{\varphi _{i - 1}}}
\end{array}} \right)$ can be constructed from the eigenvalue equation $H\left| \Phi  \right\rangle  = \varepsilon \left| \Phi  \right\rangle $ in the single-particle representation: $\left| \Phi  \right\rangle  = {\left( {\begin{array}{*{20}{c}}
  {{\varphi _1}}& \cdots &{{\varphi _N}}
\end{array}} \right)^T}$,where ${\varphi _i} = {\left( {\begin{array}{*{20}{c}}
  { a_i}&{b_i}
\end{array}} \right)^T}.$
where,
\[{L_i} = \left( {\begin{array}{*{20}{c}}
  {{\text{i}}\left( {B - Y} \right)}&{{\text{i}}\left( {A + X} \right)}&0&0 \\
  {{\text{i}}\left( {A - X} \right)}&{ - {\text{i}}\left( {B + Y} \right)}&0&0 \\
  0&0&1&0 \\
  0&0&0&1
\end{array}} \right),{R_i} = \left( {\begin{array}{*{20}{c}}
  { - \varepsilon }&{0}&{{\text{i}}\left( {B + Y} \right)}&{{\text{i}}\left( {A - X} \right)} \\
  { 0}&{ - \varepsilon }&{{\text{i}}\left( {A + X} \right)}&{{\text{i}}\left( { - B + Y} \right)} \\
  1&0&0&0 \\
  0&1&0&0
\end{array}} \right).\]
If $\textrm{Det}(L_i)\neq0$, we can get a transfer equation: $\left( {\begin{array}{*{20}{c}}
  {{\varphi _{i + 1}}} \\
  {{\varphi _i}}
\end{array}} \right) = L_i^{ - 1}{R_i}\left( {\begin{array}{*{20}{c}}
  {{\varphi _i}} \\
  {{\varphi _{i - 1}}}
\end{array}} \right) = {T_i\left( {\varepsilon} \right)}\left( {\begin{array}{*{20}{c}}
  {{\varphi _i}} \\
  {{\varphi _{i - 1}}}
\end{array}} \right)$.

Given a boundary condition, we can solve it for the physical solution, which is the function of an excitation mode. In the paper, we use the open boundary conditions ${\varphi _0} ={\varphi _{N+ 1}} = 0$ and arrive at,
\[\left( {\begin{array}{*{20}{c}}
  0 \\
  {{\varphi _k}}
\end{array}} \right) = \prod\limits_{i = 1}^N {{T_i\left( {\varepsilon} \right)}} \left( {\begin{array}{*{20}{c}}
  {{\varphi _1}} \\
  0
\end{array}} \right) \Rightarrow {\left[ {\prod\limits_{i = 1}^N {{T_i\left( {\varepsilon} \right)}} } \right]_{11}}{\varphi _1} = 0.\]
($ \left[  M\right]_{11}$ means the upper left $2\times2$ matrix.)
The necessary condition that it has physical solutions is $\textrm{Det}\left[ {{{\left[ {\prod\limits_{i = 1}^N {{T_i\left( {\varepsilon} \right)}} } \right]}_{11}}} \right] = 0$. Therefore we can get all possible energy level of excitations from it. Moreover, since the Hamiltonian $H$ is antisymmetric, the eigenvalues appear in pairs $\varepsilon_i, -\varepsilon_i$ with eigenvectors $\upsilon_i, \upsilon_i^*$, respectively. So if there is a pair of zero modes, they are Majorana fermions with the eigenvectors ${1 \over 2} (\upsilon_0+\upsilon_0^* ),  {{\rm{i}} \over 2}(\upsilon_0-\upsilon_0^* )$. It amounts to investigate the determinant $\textrm{Det}\left[ {{{\left[ {\prod\limits_{i = 1}^{N} {{T_i(\varepsilon  = 0)}} } \right]}_{11}}} \right]$. When $\varepsilon=0$,
\[{T_i} = \left( {\begin{array}{*{20}{c}}
  0&{{\mathcal{T}_i}} \\
  1&0
\end{array}} \right),{\mathcal{T}_i} = \frac{1}{{{A^2} + {B^2} - {X^2} - {Y^2}}}\left( {\begin{array}{*{20}{c}}
  {{{\left( {A + X} \right)}^2} + {{\left( {B + Y} \right)}^2}}&{2\left( {AY - BX} \right)} \\
  {2\left( {AY - BX} \right)}&{{{\left( {A - X} \right)}^2} + {{\left( {B - Y} \right)}^2}}
\end{array}} \right)\]
\[{\left[ {\prod\limits_{i = 1}^N {{T_i\left( {\varepsilon} \right)}} } \right]_{11}} = \left\{ {\begin{array}{*{20}{c}}
  0&{N = odd} \\
  {\prod\limits_{i = 1}^{N/2} {{\mathcal{T}_{2i}}} }&{N = even}
\end{array}} \right.\]
where $\textrm{Det}[\mathcal{T}_i]=1$. Thus, we can get the $Z_2$ invariant $Q=0$ for odd number of sites and $Q=1$ for even, as presented in Eq.~(8) in the paper.
Then, the pair of Majoranas read,
\begin{eqnarray*}
  {  \gamma _1} = \sum\limits_{0 \le i < N/2} {{P^{-i}}{{  \mu }_{2i + 1}}},{  \gamma _2} = \sum\limits_{0 \le i < N/2} {{P^{  i}}{{  \upsilon }_{2i + 1}}},
\end{eqnarray*}
where
\begin{eqnarray*}
  P &=& {{{A^2} + {B^2} + {X^2} + {Y^2} + 2\sqrt {\left( {{A^2} + {B^2}} \right)\left( {{X^2} + {Y^2}} \right)} } \over {{A^2} + {B^2} - {X^2} - {Y^2}}}>1, \\
  {{  \mu }_i} &=& {{ - AX - BY + \sqrt {\left( {{A^2} + {B^2}} \right)\left( {{X^2} + {Y^2}} \right)} } \over {BX - AY}}{{  a}_i} + {{  b}_i}, \\
  {{  \upsilon }_i} &=&  {{ - AX - BY + \sqrt {\left( {{A^2} + {B^2}} \right)\left( {{X^2} + {Y^2}} \right)} } \over {BX - AY}}{{  b}_i}- {{  a}_i}.
\end{eqnarray*}

When the domain wall moves to the $W$th site, $A$ and $B$ at the site turn to be $-A$ and $-B$. So, the pair of Majoranas is,
\begin{eqnarray*}
  {  \gamma _1}(W) &=& \sum\limits_{0 \le i < W/2} {{P^i}{{  \mu }_{2i + 1}}}  + \sum\limits_{W/2 \le i < N/2} {{P^{W - i - 1}}{{  \mu }_{2i + 1}}},  \\
  {  \gamma _2}(W) &=& \sum\limits_{0 \le i < W/2} {{P^{ - i}}{{  \upsilon }_{2i + 1}}}  + \sum\limits_{W/2 \le i < N/2} {{P^{ - W + i + 1}}{{  \upsilon }_{2i + 1}}}.
\end{eqnarray*}
However, if ${{A^2} + {B^2} - {X^2} - {Y^2}}=0$, there are four Majoranas in total, which are:
\begin{eqnarray*}
  {\gamma _1} = \left( {B + Y} \right){a_1} + \left( {A - X} \right){b_1}&,&{\gamma _2} = \left( {B + Y} \right){a_N} + \left( {A - X} \right){b_N},  \\
  {\gamma _3} = \left( {B - Y} \right){a_W} + \left( {A + X} \right){b_W}&,&{\gamma _4} = \left( {B - Y} \right)\left( {{a_{W - 1}} - {a_{W + 1}}} \right) + \left( {A + X} \right)\left( {{b_{W - 1}} - {b_{W + 1}}} \right).
\end{eqnarray*}
It shows that each of four Majoranas concentrates to almost one site. In this case, we can hardly distinguish which are the two Majoranas we are braiding from the other two produced by the domain wall. So, the braiding breaks down.

\section{Appendix C: Realization of $U(1)$ gauge fields}\label{B}
As mentioned in the paper, $R\left( {\alpha ,\beta } \right) = \exp \left( {{\rm{i}}{{\hat L}_z}\alpha  + 2{\rm{i}}{{\hat J}_z}\beta } \right)$ will introduce gauge fields $(\alpha+\beta,\alpha+\beta,\beta)$ into the Kramers doublet. It is known that the isospin state without gauge fields is written as:
\begin{equation}
  \left|  +  \right\rangle  = {1 \over {\sqrt 3 }}\left( {\left| {0, \uparrow } \right\rangle  + \sqrt 2 \left| {1, \downarrow } \right\rangle } \right), \left|  -  \right\rangle  = {1 \over {\sqrt 3 }}\left( {\left| {0, \downarrow } \right\rangle  - \sqrt 2 \left| { - 1, \uparrow } \right\rangle } \right),
  \label{eqS3}
\end{equation}
where $1,0,-1$ are eigenvalues of the $z$ component of the orbital angular momentum. It is easy to get:
\begin{eqnarray*}
  R\left( {\alpha ,\beta } \right)\left|  +  \right\rangle  &=& {1 \over {\sqrt 3 }}{e^{{\rm{i}}\beta }}\left( {\left| {0, \uparrow } \right\rangle  + \sqrt 2 {e^{{\rm{i}}\alpha }}\left| {1, \downarrow } \right\rangle } \right),\\
   R\left( {\alpha ,\beta } \right)\left|  -  \right\rangle  &=& {1 \over {\sqrt 3 }}{e^{{\rm{i}}\beta }}\left( {\left| {0, \downarrow } \right\rangle  - \sqrt 2 {e^{{\rm{i}}\alpha }}\left| { - 1, \uparrow } \right\rangle } \right).
\end{eqnarray*}


The rotation $R\left( {\alpha ,0} \right)$ can be realized by rotating the crystal field. On the other hand, $R\left( {0,\beta } \right)$ is proposed to be realized by an external magnetic field.
Let us consider an on-site SOC Hamiltonian with an external magnetic field along $z$ direction $H\left( t \right) = \vec l \cdot \vec S +1+ \left( {{l_z} + 2{S_z}} \right){B_z\left( t \right)}$. In the adiabatic approximation, the system remains in its instantaneous eigenstate. Supposing $B_z \left( 0 \right) = B_z \left( T \right) = 0$,  we can solve the Hamiltonian with the coupled representation and obtain,
\begin{eqnarray*}
  Eigenvalue &:& Eigenstate \\
  {E_ + }={{3 + 2B_z\left( t \right) - \sqrt {9 + 4B_z\left( t \right) + 4B_z{{\left( t \right)}^2}} } \over 4} &:& {{ - 9 - 2B_z\left( t \right) + 3\sqrt {9 + 4B_z\left( t \right) + 4B_z{{\left( t \right)}^2}} } \over {4\sqrt 2 B_z\left( t \right)}}\left| {{3 \over 2},{1 \over 2}} \right\rangle  + \left| {{1 \over 2},{1 \over 2}} \right\rangle  \\
  {E_ - }={{3 - 2B_z\left( t \right) - \sqrt {9 - 4B_z\left( t \right) + 4B_z{{\left( t \right)}^2}} } \over 4} &:& {{ - 9 + 2B_z\left( t \right) + 3\sqrt {9 - 4B_z\left( t \right) + 4B_z{{\left( t \right)}^2}} } \over {4\sqrt 2 B_z\left( t \right)}}\left| {{3 \over 2}, - {1 \over 2}} \right\rangle  + \left| {{1 \over 2}, - {1 \over 2}} \right\rangle
\end{eqnarray*}
 Notice that these two states degenerate and restore to the Kramers doublet when $B_z=0$. Since there is only one real parameter $B_z$ and the eigenstates are also real, the Berry connection $A_k=\rm{i}\left\langle \phi  \right|{\partial _k}\left. \phi  \right\rangle=0$. So, the Berry phase is absent in this process and there is only a dynamical phase $\exp \left( {{-\rm{i}}\int_0^T {{E_{\pm}}\left( t \right)dt} } \right)$ entering the Kramers doublet. In the case of ${B_z} \ll 1$, ${E_ \pm } \approx \pm {1 \over 3}{B_z}$, which will introduce two phases with opposite sign into $\left|  +  \right\rangle $ and $\left|  -  \right\rangle $, respectively. In fact, even if $E_+ +E_- \ne 0$, only an additional globe phase entering the Kramers doublet, which has no effect in our case. Now we have finalized the rotation operation $R\left( {\alpha ,\beta } \right)$.
\end{document}